\documentclass[twocolumn]{aastex631}

\shorttitle{}
\shortauthors{Keys et al.}

\begin{document}


\title{Small-scale bright point characteristics at high-resolution with the Daniel K. Inouye Solar Telescope}


\author[0000-0001-8556-470X]{Peter H. Keys}
\affiliation{Astrophysics Research center,
Queen's University of Belfast,
Northern Ireland, BT7 1NN, UK}

\author[0000-0001-5699-2991]{Ryan J. Campbell}
\affiliation{Astrophysics Research center,
Queen's University of Belfast,
Northern Ireland, BT7 1NN, UK}

\author[0009-0000-6521-8842]{Dylan K. J. Magill}
\affiliation{Astrophysics Research center,
Queen's University of Belfast,
Northern Ireland, BT7 1NN, UK}

\author[0009-0002-5370-8590]{Mateus A. Keating}
\affiliation{Astrophysics Research center,
Queen's University of Belfast,
Northern Ireland, BT7 1NN, UK}

\author[0000-0002-7725-6296]{Mihalis Mathioudakis}
\affiliation{Astrophysics Research center,
Queen's University of Belfast,
Northern Ireland, BT7 1NN, UK}

\author[0000-0002-9155-8039]{David B. Jess}
\affiliation{Astrophysics Research center,
Queen's University of Belfast,
Northern Ireland, BT7 1NN, UK}

\author[0000-0003-1746-3020]{Damian J. Christian}
\affiliation{ Department of Physics and Astronomy, 
California State University Northridge, 
Nordhoff St, Northridge, CA 91330, USA }

\author[0009-0009-8695-2558]{Arthur Berberyan}
\affiliation{Department of Astronomy \& Astrophysics, 
University of California San Diego, 
La Jolla, CA 92093, USA}

\author[0000-0001-5170-9747]{Samuel D. T. Grant}
\affiliation{Astrophysics Research center,
Queen's University of Belfast,
Northern Ireland, BT7 1NN, UK}

\author[0000-0002-7711-5397]{Shahin Jafarzadeh}
\affiliation{Astrophysics Research center,
Queen's University of Belfast,
Northern Ireland, BT7 1NN, UK}

\author[0000-0002-5365-7546]{Marco Stangalini}
\affiliation{Italian Space Agency (ASI), Via del Politecnico snc, 00133 Roma, Italy}

\author[0000-0003-3439-4127]{Robertus Erd{\'{e}}lyi}
\affiliation{Solar Physics \& Space Plasma Research Center (SP2RC), School of Mathematical and Physical Sciences, University of Sheffield; Hounsfield Road, Sheffield S3 7RH, UK}
\affiliation{Department of Astronomy, E{\"{o}}tv{\"{o}}s Lor{\'{a}}nd University; P{\'{a}}zm{\'{a}}ny P{\'{e}}ter s{\'{e}}t{\'{a}}ny 1/A, H-1117 Budapest, Hungary}
\affiliation{Gyula Bay Zolt\'an Solar Observatory (GSO), Hungarian Solar Physics Foundation (HSPF); Pet{\H{o}}fi t{\'{e}}r 3, H-5700 Gyula, Hungary}


\begin{abstract}
{
Bright points (BPs) are small-scale, dynamic features that are ubiquitous across the solar disc and are often associated with the underlying magnetic field. Using broadband photospheric images obtained with the Visible Broadband Imager at the National Science Foundation's Daniel K. Inouye Solar Telescope (DKIST), the properties of BPs have been analyzed with DKIST for the first time at the highest spatial resolutions achievable. BPs were observed to have an average lifetime of $95\pm29$~s and a mean transverse velocity of $1.60\pm0.41$\,km\,s$^{-1}$. The BPs had a log-normal area distribution with a peak at $2300$\,km$^2$. Transverse velocity and lifetimes across the DKIST images were comparable and consistent with previous studies. The area distribution of the DKIST data peaked in areas significantly lower than those from the literature. This was explored further and was observed to be due to an overestimation of BP areas due to the merging of close features when the spatial resolution is reduced, in tandem with possible over-splitting of features in the DKIST images. Furthermore, the effect of variable seeing within the data was determined. This showed that the average spatial resolution of the data was around $0{\,}.{\!\!}{''}034\pm0{\,}.{\!\!}{''}007$ in comparison to the theoretical diffraction-limit of $0{\,}.{\!\!}{''}022$. Accounting for the influence of seeing, the peak of the area distribution of BPs in the DKIST data was estimated as $4800$\,km$^2$, which is still significantly lower than previously observed. 
} 
{}
{}
\end{abstract}
\keywords{Sun: photosphere --- Sun: magnetic fields --- Sun: granulation}


\section{Introduction} \label{sec:intro}
 Bright points (BPs) are observed throughout the solar cycle and across the entire solar disc. They are often associated with the underlying magnetic field within the intergranular lanes, and are sometimes referred to as magnetic bright points \citep{Solanki1993, deWijn2009}. The theory  of how these features form has been around for some time \citep{Roberts1978, Parker1978, Spruit1979a, Spruit1979b}. The theory posits that magnetic flux is advected into the intergranular lanes where it concentrates. Fast downflows within the intergranular lanes and pressure differences between the flux tube and the surroundings cause the magnetic flux concentrations to contract until the outward pressure due to the intensified magnetic field balances the external pressures exerted on the flux tube. The equipartition field strength at this point is expected to be on the order of a kilogauss. This process has been referred to as `convective collapse' due to the role of convective processes in enhancing the magnetic field strength. The concentration of the field makes the feature appear point-like. Then, a combination of heating at the base of the tube from the surrounding hot granular walls, and the fact that the tube is partially evacuated, means that the intensity of the features will be greater than that of their surroundings.

 The convective collapse process was eventually observationally identified with Hinode data by \citet{Nagata2008}. The authors found that there was an increased downflow (up to 6~km\,s$^{-1}$) in the region of a magnetic field concentration, which then increased the magnetic field strength to 2~kG and enhanced the intensities observed in the G-band as a BP formed. It should be noted that G-band is often used to study BPs, as it has been shown that the abundance of CH molecules is reduced at the higher temperatures experienced within the flux tube of the BP \citep{Steiner2001}. Therefore, BPs will tend to appear brighter in G-band intensity images in comparison to other photospheric bandpasses. 

Theory has suggested that magnetic fields within BPs are of the order of a kilogauss, which has been reported in both simulations and observations \citep[][to name a few]{2010ApJ...719L.134J, Utz2013, Criscuoli2014, Buehler2019, CubasArmas2021}. However, studies have shown that the magnetic field strength of BPs follows a bimodal distribution \citep{Utz2013,Keys2019}, with a `weak' group at around $300$\,--\,$600$~G and a `strong' group at around $1100$\,--$1300$~G. \citet{Utz2014} showed that only $30$\% of BPs reach kilogauss field strengths. It had been posited \citep{Keys2019} that this `weak' group was due to concentrations of magnetic field within the lanes that had yet to undergo convective collapse, with emerging flux being dispersed before it could amplify to kilogauss strengths. Subsequent work by \citet{Keys2020} showed that several processes, aside from solely convective collapse, can amplify the field strength and give rise to kilogauss field strengths. 

BPs have been estimated as covering $\sim0.9$\% -- $2.2$\% of the solar surface \citep{SanchezAlmeida2010}, with diameters typically between $\sim100$\,--\,$300$~km \citep{SanchezAlmeida2004, Utz2009, Crockett2010}. Furthermore, BPs have been observed to expand with height \citep{MartinezGonzalez2012, Kuckein2019}. Reported lifetimes can vary between 90~s to several minutes depending on the region in which they are found and the characteristics of the observational data \citep{Utz2010, Keys2011}. BPs are dynamic features with typical transverse velocities of around $1$~km\,s$^{-1}$ \citep{Utz2009, Keys2011}. BPs can experience rapid excursions above $3$~km\,s$^{-1}$ due to super diffusive L{\'e}vy flights \citep{Jafarzadeh2013}, with diffusion in the photosphere linked to the position of the BPs within the network cell. Recently, \citet{Stangalini2025} found that the dynamic behaviours of small-scale magnetic field concentrations are modulated by the solar cycle, with temporal scales on the order of $11$~years, although at much lower spatial resolutions ($1{\,}\,{\!\!}{''}$) than is commonly achievable with ground-based facilities.

BPs are often seen as important in studying the evolution of magnetic fields at small spatial scales \citep{BellotRubio2019} as they are interpreted as tracers of flux tube footpoints \citep{Yang2015}. Furthermore, their dynamic behaviour and magnetic nature means that they are frequently analyzed in wave studies as a possible source of propagating wave phenomena \citep{Jess2023, Berberyan2024}. By analyzing high-resolution simulations, \citet{Keys2021} were able to show that inversions of a BP with a known driver could return accurate atmospheric parameters at the resolution of DKIST in the presence of an upwardly propagating wave. Therefore, the analysis of BPs with DKIST could be beneficial in understanding their contributions to channeling energy to higher regions in the solar atmosphere. Here, we present the first analysis of the characteristics of BPs with DKIST. The study looks at the properties of these BPs at the highest resolution ever obtained. 
 

\section{Observations}\label{sect:observations}
\begin{figure*}
    \centering
    \includegraphics[width=\textwidth]{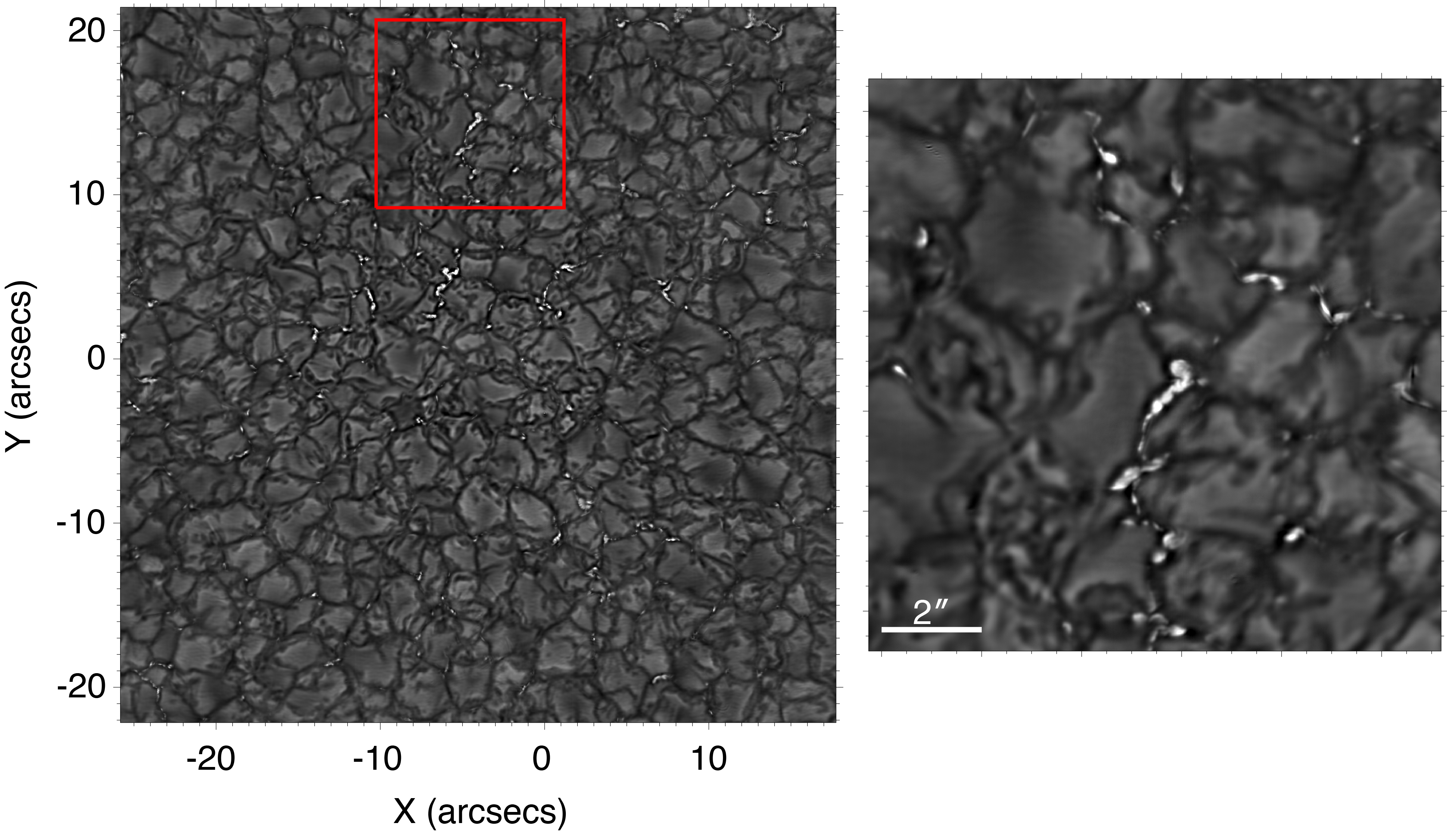}
    \caption{A sample image of the quiet Sun data obtained with VBI on 26 May 2022 with the G-band filter at 17:50UT. The images have a spatial sampling of $0{\,}.{\!\!}{''}011$~pixel$^{-1}$ and a cadence of around 6~s. The estimated spatial resolution of this image is $0{\,}.{\!\!}{''}032$ ($\sim 23$\,km). The full field-of-view of the image can be seen in the left hand panel. The red box indicates the region shown in the zoomed in  right hand panel. The zoom indicates the typical structure and configurations of the bright points under investigation at the resolution of DKIST. Examples of both elongated chains, bright point groups, as well as isolated point-like features, can be seen.}
    \label{fig_fov}
\end{figure*}

On 2022 May 26, between 17:46UT and 19:31UT, a sequence of quiet Sun observations at disc center were obtained with the Visible Broadband Imager \citep[VBI;][]{VBI2021} installed as a common user instrument at the Daniel K. Inouye Solar Telescope \cite[DKIST;][]{DKIST2020} during its first cycle of operations. The observing sequence was chosen to maximise the spatial and temporal capabilities of the instrument to analyze small-scale, short-lived features at the highest possible resolutions. The 2.8 arcminute field stop was employed with a spatial sampling of $0{\,}.{\!\!}{''}011$~pixel$^{-1}$ giving an effective field-of-view of around $45'' \times 45''$. Observations were taken with the $4305$~{\AA} G-band filter at a cadence of around $6.1$~s. A sample image of the field-of-view with a cutout showing a sample of BPs in the image is given in Figure~\ref{fig_fov}.

This observing sequence has associated co-spatial images taken with the Ca~{\sc{ii}}~K filter with VBI and scans with the Visible Spectropolarimeter \cite[ViSP;][]{ViSP2022} in the Fe~{\sc{I}} line pair at $6301.5$~{\AA} and $6302.5$~{\AA}, alongside the Ca~{\sc{ii}} $8542$~{\AA} line. Within this study, we limit our analysis to the G-band, however, other aspects of this observing sequence have been studied and detailed elsewhere \citep[e.g.,][]{Campbell2023}. Although the magnetic field can be returned with ViSP, the scan time for these observations, coupled with the expected lifetimes of any identified BPs means that it would be difficult to return magnetic field distributions for a significant sample of BPs within the data. Furthermore, the dynamic nature of BPs means that there is the potential for BPs to move outside the slit during the scan with ViSP. Studies of BPs with ViSP would then be limited to longer-lived features that are likely to be more static. Also, the spatial resolution of ViSP is approximately an order of magnitude worse than VBI, so spatially resolving BPs would be more challenging. Here, we limit the study to G-band due to the superior spatial and temporal resolution with VBI. Also, with G-band, we can more readily compare to previous work and BPs in G-band are generally considered as tracers of the magnetic field, as mentioned previously.

It should be noted that although the sequence ran for around 1~hour and 45~minutes (1034 frames in total), the location of the adaptive optics lock point shifted at around 90~minutes into the sequence due to a temporary drop in seeing conditions. Therefore, care must be taken when analyzing the continuation of features around this point in the sequence. Seeing conditions were relatively good and consistent across the data set aside from this temporary drop near the end of the sequence. The data has a mean Fried parameter (r$_0$) of $12.1\pm 3.1$~cm and a range of $10.2$~cm to $20.9$~cm as reported in the FITS metadata of the images.

The data underwent level 1 processing onsite, whereby the images had been flat and dark corrected, as well as undergoing speckle reconstruction \citep{Woger2008} prior to being made publicly accessible. A post-speckle reconstruction pipeline was generated to produce science-ready images from the level 1 data. This pipeline was based on the Rapid Oscillations in the Solar Atmosphere \citep[ROSA;][]{ROSA2010} instrument pipeline, which is a similar instrument to VBI, providing multi-channel broadband imaging. The pipeline performs image destretching to account for residual atmospheric seeing effects, as well as aligning the data to remove pointing fluctuations and co-aligning data between multiple filters for the image sequence. An example of the fully-processed data is shown in Figure~\ref{fig_fov}.


\section{Methods}\label{sect:methods}
BPs were tracked in the G-band images using an automated detection and tracking algorithm \citep{Crockett2010}, which has been employed across several studies to identify features between frames in G-band \citep{Keys2011, Keys2014} and other wideband filters \citep{Keys2019, Keys2020}. As a short summary, the code works by isolating bright features in an image before using intensity thresholding to determine the boundary locations of BPs and discarding erroneous detections, such as exploding granules or brightenings at the edge of granules due to density enhancements. Due to the scale of the images acquired from DKIST, and the way that memory allocations are handled by the algorithm, the code had to be augmented to improve memory management through parallel processing. Even with these improvements, and the use of high-performance computing to process the data, this phase took a significant number of compute hours in comparison to similar data from smaller aperture telescopes (nearly $10$ times longer than images from 1~m class telescopes). This is not unexpected, but something to consider for future work.

\begin{figure*}
    \centering
    \includegraphics[width=\textwidth]{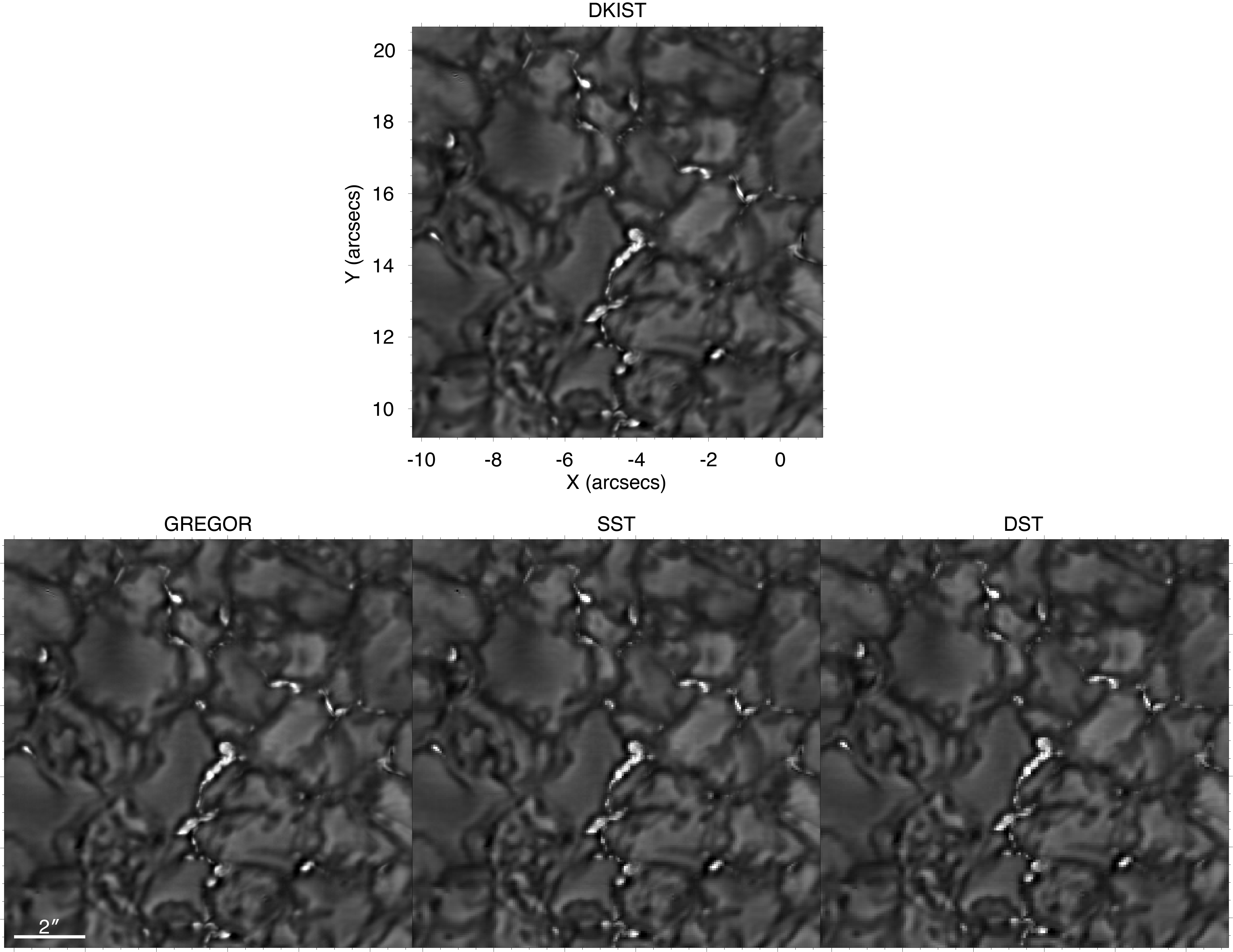}
    \caption{Samples images of the DKIST data degraded to match the resolution of other commonly used ground-based facilities (GREGOR, SST and DST). Degradation was performed using a similar methodology as outlined in \citet{Campbell2021}. The top panel shows the zoomed image from Figure~\ref{fig_fov} of the original DKIST image. The labeled panels below show, from left to right, the DKIST data degraded to the equivalent resolution of GREGOR, the SST and the DST, respectively. This figure shows the effect of spatial resolution on the observed structure of BPs for both BP chains and isolated features.}
    \label{degraded_inten_sample}
\end{figure*}

To compare the DKIST VBI image sequence to similar data from other high-resolution ground-based facilities, the DKIST data was degraded to match the spatial sampling of typical data from GREGOR \citep{GREGOR2020}, the Swedish Solar Telescope \citep[SST;][]{SST2003} and the Dunn Solar Telescope (DST). The approximate spatial sampling for the degraded data sets were $0{\,}.{\!\!}{''}038$~pixel$^{-1}$, $0{\,}.{\!\!}{''}059$~pixel$^{-1}$, and $0{\,}.{\!\!}{''}069$~pixel$^{-1}$ for GREGOR, SST, and DST data, respectively. The spatial degradation was performed using an approach similar to that outlined by \citet{Campbell2021}, with the DKIST observations convolved with theoretical PSFs of the respective facilities. The temporal resolution of the degraded data was kept the same as that of the original DKIST data, as each of these facilities can easily acquire data with a 6~s cadence. The degraded data sets were analyzed with the same detection and tracking algorithm as the DKIST data. A sample of the degraded images detailing a zoomed in example of a BP at these resolutions is included in Figure~\ref{degraded_inten_sample}. For simplicity, the degraded data will henceforth be referred to in terms of the facility the data was degraded to match (i.e., GREGOR, SST and DST), while the original data will be referred to as the DKIST data.

Furthermore, to judge the effect of seeing on the BP properties established, the spatial resolution for each frame of the DKIST data was estimated using the Fourier technique outlined in Appendix A of \citet{Beck2007}. Figure~\ref{spat_res_var} shows the variation of spatial resolution over time for this particular data set. The mean spatial resolution across the data set was found to be $0{\,}.{\!\!}{''}034\pm0{\,}.{\!\!}{''}007$, with the spatial resolution varying from around $0{\,}.{\!\!}{''}023$ (close to the theoretical diffraction limit) to $0{\,}.{\!\!}{''}066$. It should be noted here that the average spatial resolution at this wavelength would correspond to the diffraction limit of a 3-m class telescope. An estimate of r$_0$ is given within the image metadata. However, since the value of r$_0$ depends on where in the Earth's atmosphere the value is estimated, the spatial resolution calculated for each frame via the Fourier approach will give a better indication of image quality. It should of course be noted here that this technique is also not without limitations. The technique involves an estimate of the frequency cutoff to determine the spatial resolution. This choice can be somewhat subjective and it is possible that some signal may be retained above the identified cutoff, which would result in a lower estimate of the spatial resolution than reported. However, the technique is still likely to be a more reliable estimate of image quality over r$_0$. For part of the analysis, the spatial resolution in each frame was used to determine the smallest resolvable pixel dimensions for BPs in each frame to filter the subsequent area and velocity estimates of the BPs. Both distributions are discussed below.

\begin{figure}
    \centering
    \includegraphics[width=\columnwidth]{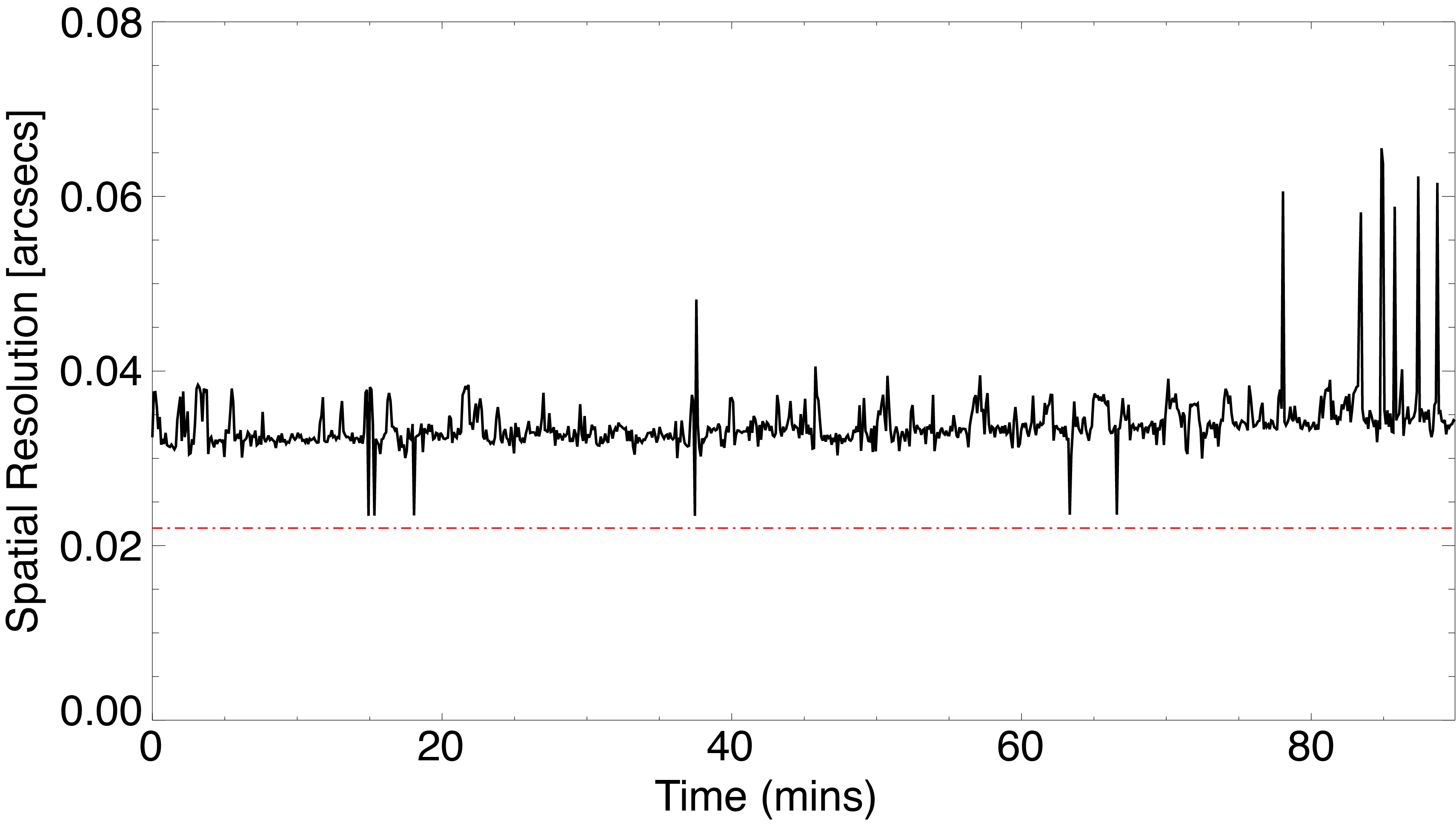}
    \caption{The variation of spatial resolution over the course of observations. The spatial resolution is estimated with Fourier techniques outlined in \citet{Beck2007}. The zero time is the first frame of the observations obtained at 17:46UT on 2022 May 26. The plot is limited to images prior to an AO lock point jump, which occurs at about 90~minutes into the sequence. The drop in image quality prior to the loss of the AO lock is evident in the plot. The {\textit{red dot-dashed}} line indicates the theoretical diffraction-limited resolution of the data acquired. The average spatial resolution for the sequence is $0{\,}.{\!\!}{''}034\pm0{\,}.{\!\!}{''}007$.}
    \label{spat_res_var}
\end{figure}

With the transverse velocities, the diffusive properties of the BPs were estimated. Several authors have previously looked at the diffusive properties of BPs \citep{Cadavid1999, Abramenko2011, Jafarzadeh2014}. Here, the diffusion index, $\gamma$, was estimated by analyzing the squared-displacement for each BP over time. With this, the diffusion coefficient \citep{Monin1975}, $K$, is calculated using the diffusion index. The diffusion coefficient indicates the efficiency of BP dispersal, therefore, these metrics are useful to understand BP dynamics and field dispersal at these scales.


\section{Results}\label{sect:results}

\subsection{Bright Point Area}\label{sect:areas}
Across the DKIST data, 12{\,}486 individual BPs were identified. Adjusting for differences in the field-of-view and the duration of the observations between this data and the data used in \citet{Keys2011} (DST data analyzed with the same algorithm), $\sim 2.7$ times more BPs were identified within this data set. The increase in numbers detected is due to a greater number of smaller features identified as a result of the superior resolving power of DKIST.

\begin{figure}
    \centering
    \includegraphics[width=\columnwidth]{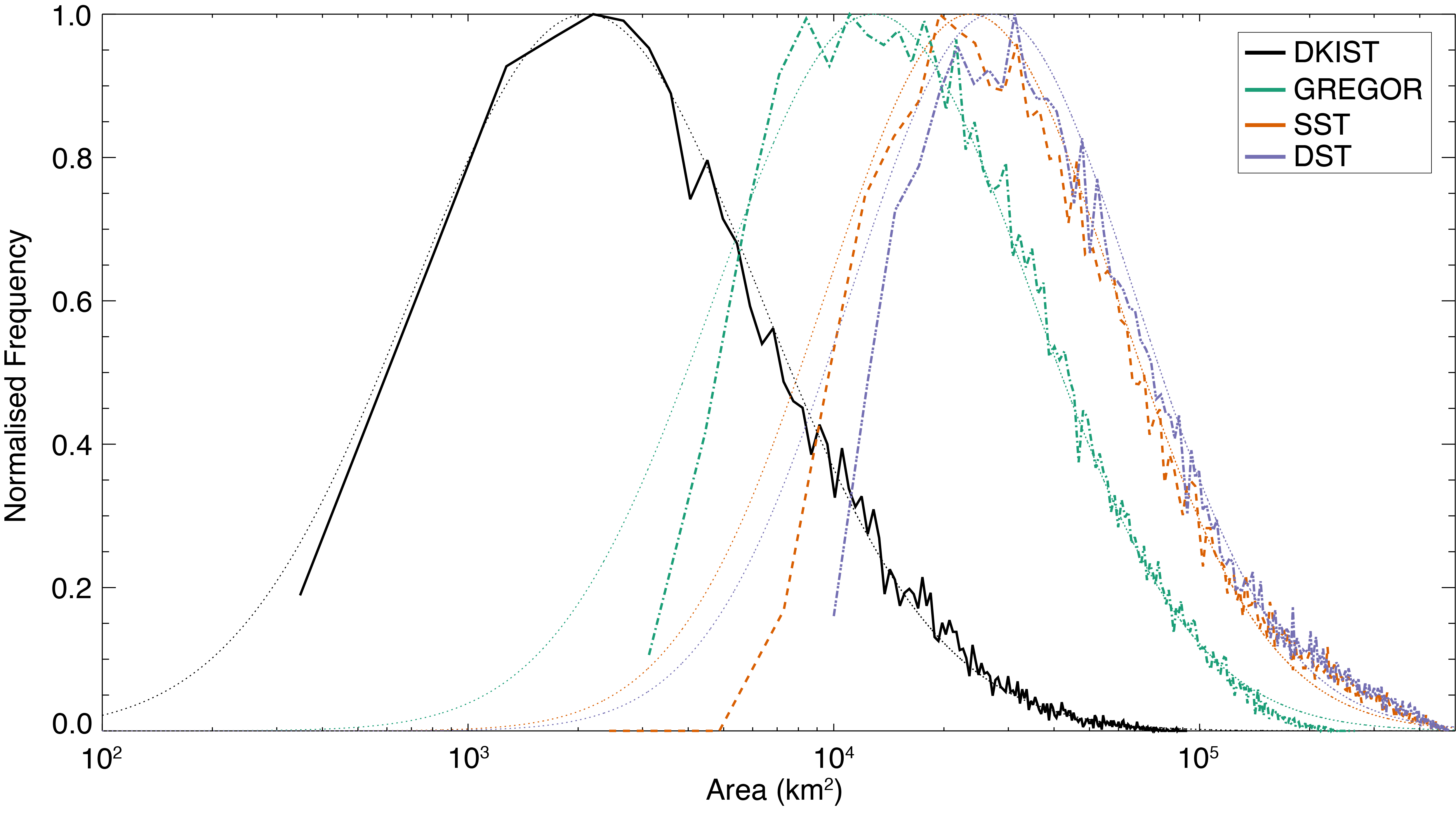}
    \caption{The area distribution of BPs identified in the datasets. The {{\textit{solid}}} {\textit{black}} line represents the original DKIST data, while the {{\textit{dot-dashed green}}}, {{\textit{dashed orange}}}, and {{\textit{dot-dot-dashed blue}}} lines represent the area distributions for data degraded to the resolution of GREGOR, SST and DST, respectively. The {{\textit{dotted}}} lines indicate the corresponding log-normal fits for each of the distributions. The peak of the distributions occurs at approximately $2300$\,km$^2$, $13{\,}700$\,km$^2$, $25{\,}700$\,km$^2$ and $26{\,}200$\,km$^2$ for the DKIST, GREGOR, SST, and DST data, respectively.}
    \label{degraded_area_dist}
\end{figure}

The area distribution of the BPs identified is shown in Figure~\ref{degraded_area_dist} (\textit{black} line). The peak of the distribution corresponds to an area of $\sim 2300$\,km$^2$, which would give an effective diameter of a BP (assuming a circular geometry) of 54~km. It should be clear from Figure~\ref{fig_fov} that BPs do not typically have a circular geometry, with extended chains being somewhat common and smaller BPs resembling more of a `raindrop' shape. However, we report the equivalent diameter here, as it is a metric that is often reported in the literature with regards to BP areas \citep[e.g.,][]{Utz2009}. A study of high resolution MHD simulations by \citet{Peck2019} find BPs range from circular shapes to elongated sheets with most structures having a width of $\sim 70$\, km.

The peak in area here is substantially lower to those reported previously. To explore the reasons behind this further, as described above, the spatial resolution of the data was degraded to match that of three other facilities/instruments (GREGOR, SST, and DST), while the temporal resolution was kept the same. The same process of tracking the data was performed on the degraded data identifying 3382, 2937 and 2849 BPs in the GREGOR, SST, and DST data respectively. The purpose of this was to ascertain how the area distribution and characteristics varied across the different facilities to confirm whether the variations are solely due to the resolving power of DKIST, or if there is another reason explaining the discrepancy. By degrading the existing DKIST data, we would then be able to have a like-to-like comparison for identified features in the images. That is, the same BPs should be contained within the data; it will only be the change in spatial resolution that would result in any changes to the BP characteristics identified.

Figure~\ref{degraded_area_dist} shows the effect of the spatial resolution on the BP area distributions. The {\textit{dot-dashed green}}, {\textit{dashed orange}}, and {\textit{dot-dot-dashed blue}} lines show the corresponding distributions for the data degraded to the resolutions of GREGOR, the SST, and the DST, respectively, while the {\textit{dotted}} lines correspond to the log-normal fits for the associated data. The peak for the distributions occur at approximately $13{\,}700$\,km$^2$, $25{\,}700$\,km$^2$ and $26{\,}200$\,km$^2$ for the GREGOR, SST, and DST data, respectively. In a circular geometry this corresponds to a diameter of around $132$\,km, $181$\,km, and $183$\,km, for the GREGOR, SST, and DST data, respectively. 

\subsection{Varying Spatial Resolution}\label{sect:res}

\begin{figure}
    \centering
    \includegraphics[width=\columnwidth]{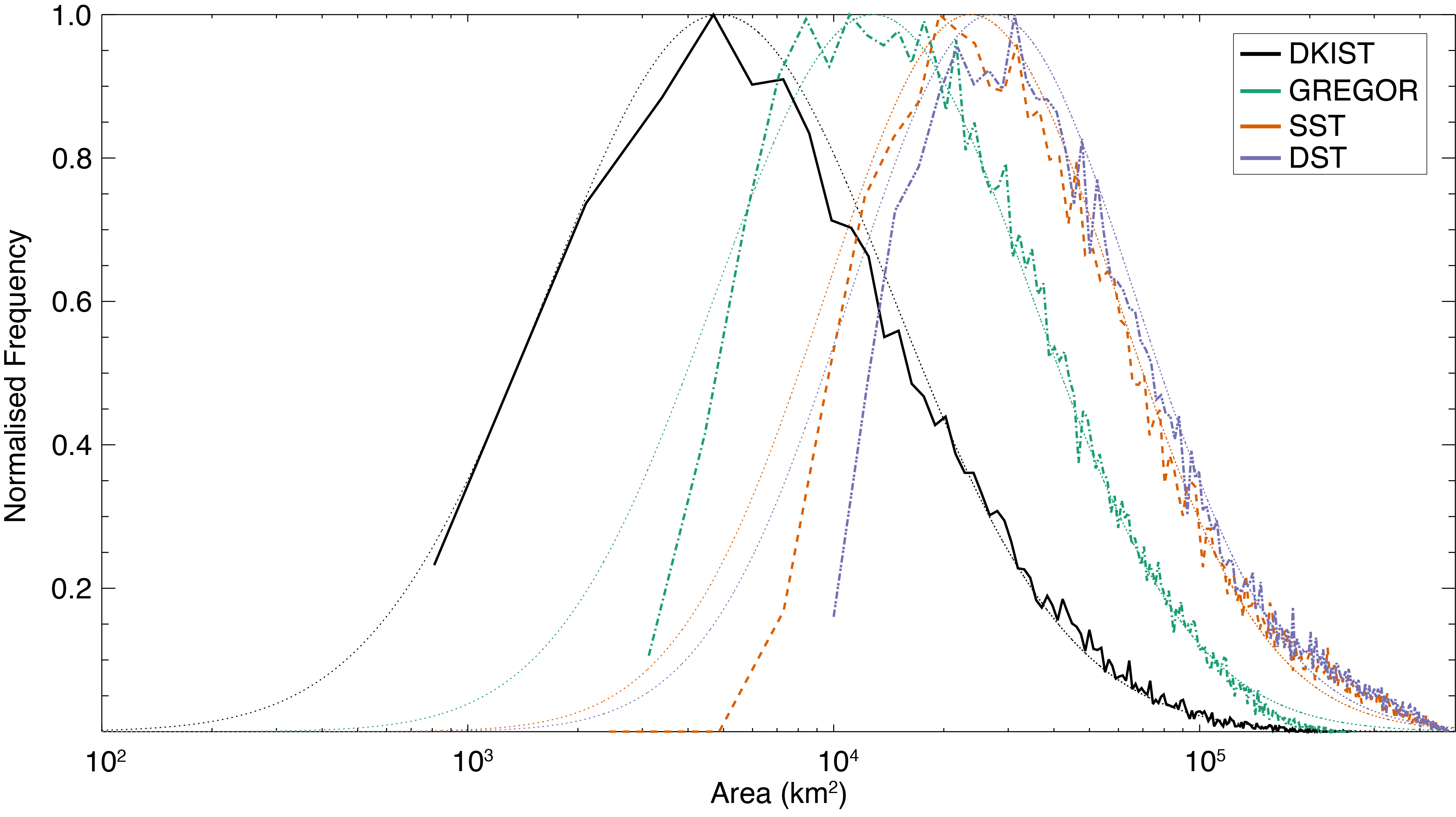}
    \caption{The area distribution of DKIST BPs adjusted for variations in seeing across the data set. The filtered DKIST distribution is displayed with the {{\textit{solid}}} {\textit{black}} line while the associated DKIST data degraded to the resolution of other facilities, as shown in Figure~\ref{degraded_area_dist}, are included here again for reference. Again, the {{\textit{dotted}}} lines show the associated log-normal fits for the distributions. Accounting for the variable seeing conditions across the data set shifts the peak location to around $4800$\,km$^2$. The peak of the distribution occurs at smaller areas than those of the degraded data sets; however, accounting for seeing returns a distribution shape that matches closer to those previously reported in the literature for facilities with reduced spatial resolution. The effects of seeing are pronounced with the DKIST data, so care must be taken in analyzing results obtained with these data.}
    \label{corrected_area}
\end{figure}

Having estimated the spatial resolution in each frame (see Section~\ref{sect:methods}), this value was then used to filter area estimates in each frame, so that only BPs that are resolvable in a given frame are considered. To do this, the spatial resolution estimate in a given frame was used to estimate the smallest resolvable pixel dimensions for a BP within that frame. Any BP that fell below that limit in that frame was then removed from the sample, as it was now deemed unresolvable due to seeing, regardless of whether the algorithm was able to identify a feature of interest previously. This meant that seeing variations could be accounted for when determining the area distribution of the BPs. The resulting distributions can be seen in Figure~\ref{corrected_area}. The peak of this distribution is around $4800$\,km$^2$, which corresponds to a diameter of 78\,km when assuming a circular geometry. This is still significantly lower than previously reported area distributions. It should also be reiterated that the technique used to estimate the spatial resolution has the potential to overestimate the spatial resolution.

Extending this analysis further, the spatial resolution estimates were used to isolate the area distributions for the best frames. As shown in Figure~\ref{spat_res_var}, there were 6 frames with a resolution better than $0{\,}.{\!\!}{''}026$. From these 6 frames 828 BPs were identified. The area distribution of BPs from these best frames was seen to match closely with the DKIST area distribution shown in Figure~\ref{degraded_area_dist}, with a peak at $~2100$\,km$^2$.

\subsection{Bright Point Dynamics}\label{sect:vels}
\begin{figure}
    \centering
    \includegraphics[width=\columnwidth]{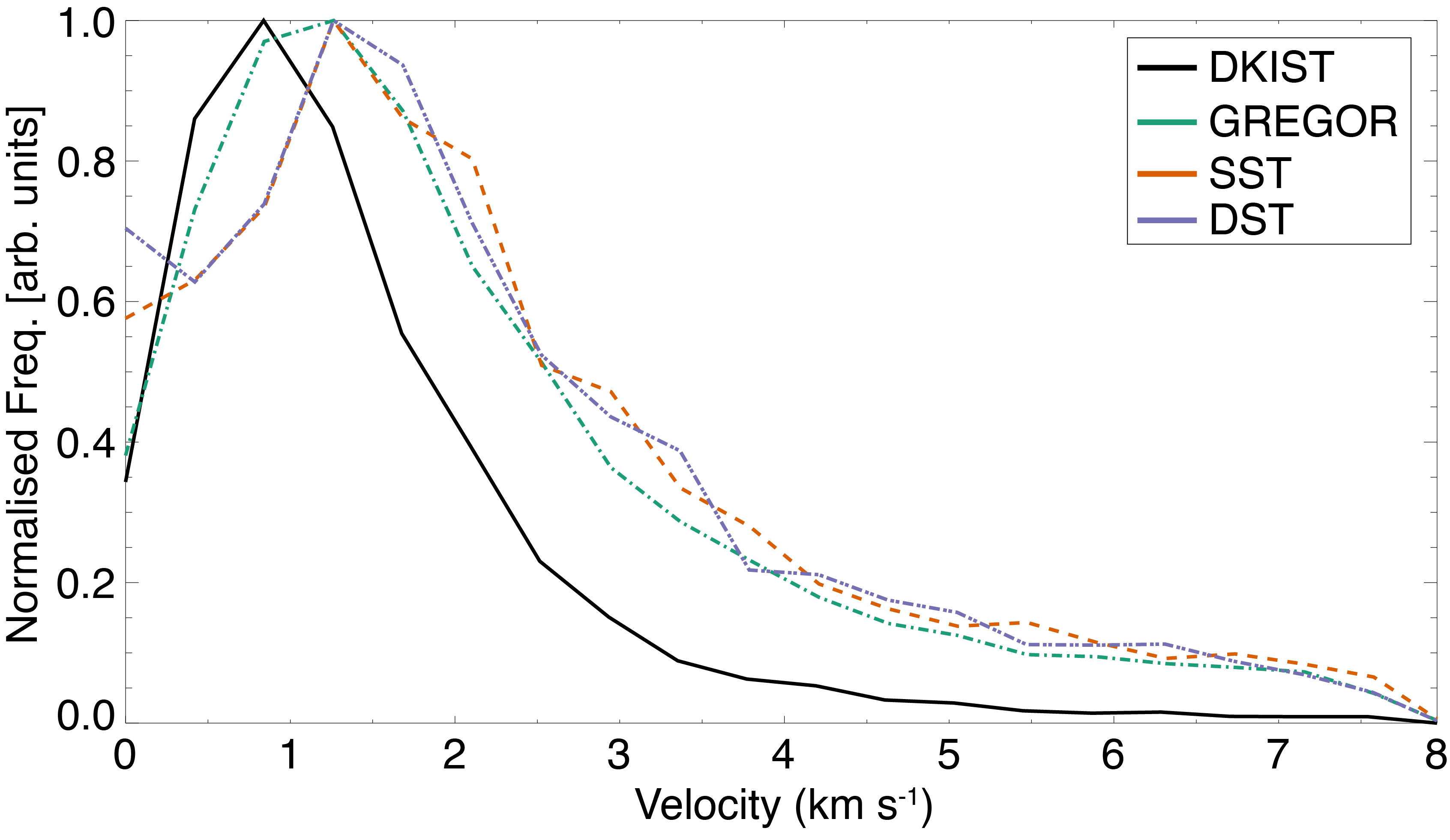}
    \caption{The velocity distribution of BPs identified in DKIST data and the DKIST data degraded to the resolution of GREGOR, the SST and the DST. The {{\textit{solid}}} {\textit{black}} line represents the original DKIST data, while the {{\textit{dot-dashed green}}}, {{\textit{dashed orange}}}, and {{\textit{dot-dot-dashed blue}}} lines represent the area distributions for data degraded to the resolution of GREGOR, SST and DST, respectively. The peak of the distributions occur at approximately $0.83$\,km\,s$^{-1}$ for DKIST and $1.25$\,km\,s$^{-1}$ for the degraded data sets. The mean BP velocity is $1.60\pm0.41$\,km\,s$^{-1}$, $2.30\pm0.64$\,km\,s$^{-1}$, $2.44\pm0.68$\,km\,s$^{-1}$, and $2.36\pm0.66$\,km\,s$^{-1}$ for the DKIST, GREGOR, SST and DST data, respectively.}
    \label{Fig_vel_char}
\end{figure}

As well as establishing the area of the BPs, the tracking algorithm recorded the location of the BPs in each frame in which they were identified. This allowed analysis of the dynamic properties of BPs within the data (i.e., transverse velocities, lifetimes, and diffusion characteristics). Again, this was performed for both the DKIST data and the data degraded to the resolutions of GREGOR, the SST, and the DST. The distributions of velocities across the data sets are given in Figure~\ref{Fig_vel_char}.

The distribution of velocities for the DKIST data appears to have a sharper peak and a narrower tail, with few BPs displaying higher velocities. A greater number of higher velocity BPs are detected in the degraded data sets. The peak of the distribution for the DKIST data occurs at approximately $0.83$\,km\,s$^{-1}$. The peak for the degraded data sets appears at roughly the same velocity of $1.25$\,km\,s$^{-1}$, although the GREGOR data has a slightly broader peak than the SST and the DST data sets. The shape of the distributions somewhat mirrors those of the area distributions. The results across all data sets are consistent with previous studies of BP velocity distributions \citep{Keys2011}. 

\begin{figure}
    \centering
    \includegraphics[width=\columnwidth]{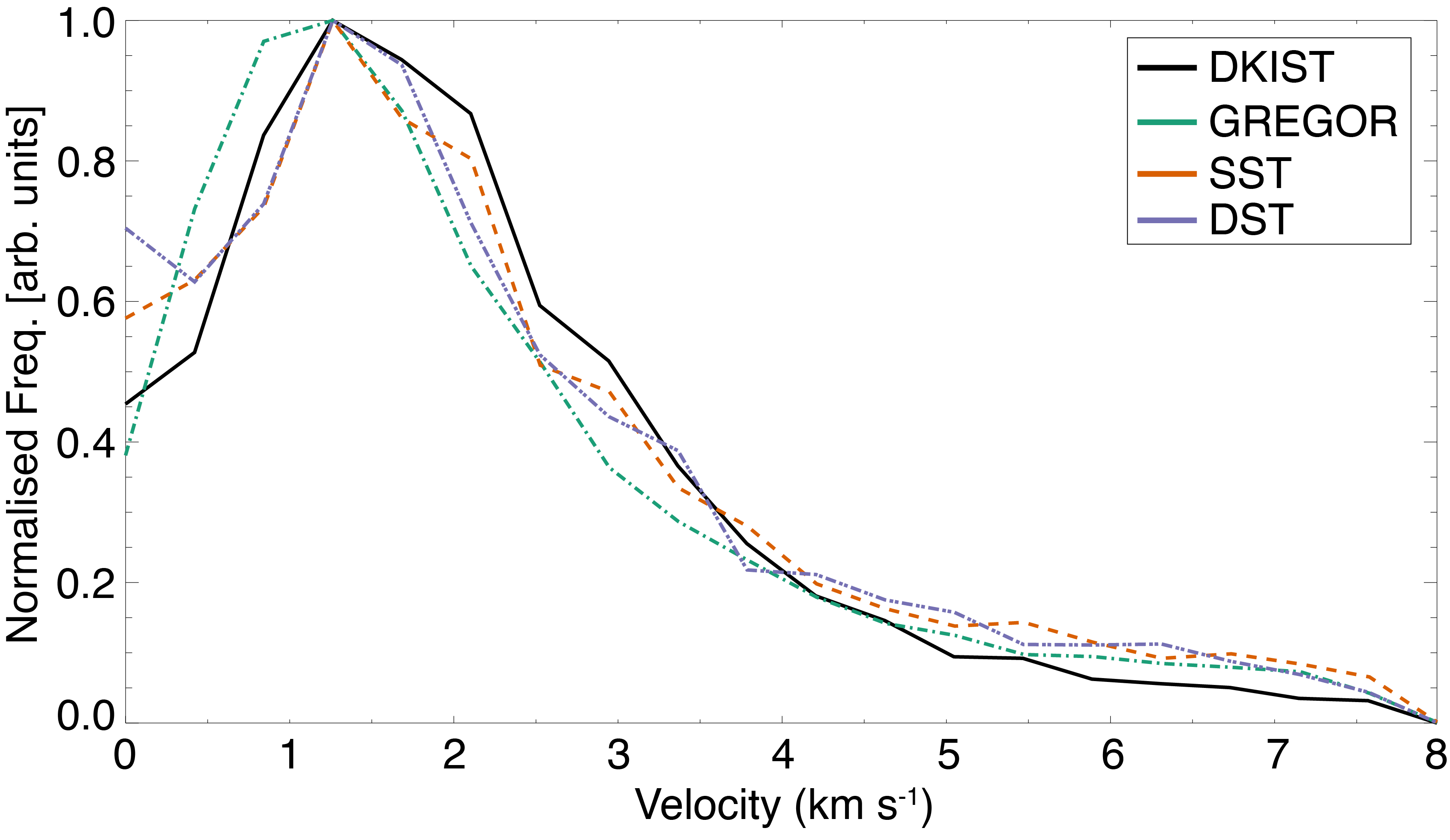}
    \caption{The velocity distribution of BPs identified in DKIST data adjusted for variations in seeing across the data set. The filtered DKIST distribution is displayed with the {{\textit{solid}}} {\textit{black}} line while the associated DKIST data degraded to the resolution of other facilities, as shown in Figure~\ref{Fig_vel_char}, are included here again for reference. Accounting for fluctuations in seeing broadens the distribution and shifts the peak so that the DKIST velocity distributions match closer to those of the degraded data sets. The mean velocity for the filtered DKIST data is $2.28 \pm 0.58$~km\,s$^{-1}$.}
    \label{corrected_vel}
\end{figure}

Similar to the issues discussed in Section~\ref{sect:res}, variable spatial resolution will have an effect on the estimated velocities as the capabilities to resolve BPs will vary with the change in resolving power. Again, the variable spatial resolution is accounted for in the velocity estimates for the BPs in a manner similar to that outlined in Section~\ref{sect:res}. The results are shown in Figure~\ref{corrected_vel}. Accounting for the spatial resolution in each frame broadens the velocity distribution of the DKIST BPs and moves the peak of the distribution to a slightly larger velocity value. The distribution now matches the distributions for the degraded data, particularly with regard to the shape of the distribution. The mean BP velocity for the filtered data is found to be $2.28 \pm 0.58$~km\,s$^{-1}$, which again is similar to the degraded data sets. 

The diffusion index, $\gamma$, is a useful metric for determining whether the BP motions follow normal diffusion ($\gamma = 1$), or are super-diffusive ($\gamma > 1$)  or sub-diffusive ($\gamma < 1$). The value for $\gamma$ for the DKIST data was $1.22\pm0.41$, while the value is $1.24\pm0.43$ when accounting for seeing using the spatial resolution estimates of the frames. For the degraded data sets $\gamma$ is estimated as $1.40\pm0.58$, $1.52\pm0.69$, $1.55\pm0.73$ for the GREGOR, SST, and DST data, respectively. This is slightly higher than the DKIST results, but still within the associated errors of one another, and implies that the BPs follow a super-diffusive regime. 

The diffusion coefficient, $K$, indicates the efficiency of the BPs dispersal. For the DKIST data, $K$ was found to be $30\pm18$~km$^2$\,s$^{-1}$ and $74\pm47$~km$^2$\,s$^{-1}$ when the spatial resolution filtering is applied. For the degraded data sets, $K$ is estimated as $98\pm55$~km$^2$\,s$^{-1}$, $142\pm83$~km$^2$\,s$^{-1}$, and $144\pm83$~km$^2$\,s$^{-1}$ for the GREGOR, SST, and DST data, respectively. 

It should be noted that the estimated average lifetime of the BPs was established to be $95\pm29$~s, $97\pm32$~s, $98\pm33$~s, and $97\pm33$~s for the DKIST, GREGOR, SST, and DST data, respectively. The values for the lifetimes are in agreement across all the data sets.

More prominent variations appear when considering the longest-lived BPs detected. For the DKIST data, the maximum BP lifetime was found to be $\sim 22$~minutes, whereas for the GREGOR, SST, and DST data sets, the longest lifetimes were $\sim 35$~minutes. These do not correspond to the same features, though they correspond to BPs within the same spatial region (i.e., around $-10''$ to $0''$ in $x$ and $0''$ to $10''$ in $y$ in Figure~\ref{fig_fov}). The average velocity across their lifetime for these longer lived BPs are below $0.5$\,km\,s$^{-1}$ for all four data sets, suggesting that longer-lived BPs are less dynamic. Similar behavior with regards to lifetime and velocity has been reported in the literature \citep{Chitta2012, Keys2014, BerriosSaavedra2022}. It should be noted that degradation was performed on the spatial dimensions and not the temporal scale. Therefore, any differences with respect to lifetimes are due to the classification of BPs at lower spatial resolutions by the code, as opposed to being the result of poorer temporal sampling.


\section{Discussion}\label{sect:discussion}

\subsection{Bright Point areas}\label{sect:disareas}

The area distribution is fit well with a log-normal distribution as seen in Figure~\ref{degraded_area_dist} ({\textit{dotted black} line)}. Such distributions have been reported previously for BPs \citep{Crockett2009, Keys2014, BerriosSaavedra2022}, so this is not unexpected. However, the distribution for the area in this DKIST data has a peak at a lower area ($\sim 2300$\,km$^2$) and has a sharper tail than previous observations \citep[e.g., a mean peak in area of $27,000$\,km$^2$ as reported by][]{Keys2014}. The change in location of the peak is perhaps not unexpected. The theoretical diffraction-limited resolution is substantially better with DKIST than with any other facility. This would indicate that the average BP detected in DKIST data is smaller than at other facilities and that smaller features were perhaps not resolved adequately previously. Furthermore, a sharper tail in the distribution would suggest that there are fewer longer chains of BPs than previously observed. Again, this is perhaps not unexpected, as the superior spatial resolution afforded by DKIST may be able to distinguish separate entities in larger groups of BPs, which perhaps were unresolved previously.

The degradation of the data to lower spatial resolutions, as expected, resulted in changes in average sizes of the BPs. The distributions still follow log-normal distributions quite well, however, the peak of these distributions occurs at larger area values ($13{\,}700$\,km$^2$, $25{\,}700$\,km$^2$, and $26{\,}200$\,km$^2$ for the GREGOR, SST, and DST data, respectively), which matches well to the lower spatial resolutions of the facilities. That is, at lower resolutions, the peak of the distribution is pushed to higher values. In addition, the tail of the distribution is broader for these degraded images. 

These results are not unexpected, however, the difference in the area peak values obtained with each data set does not correspond solely due to the change in scaling of the spatial resolution. That is, the shift in the location of the peak is not proportional to the variation in pixel scale. Therefore, with the change in spatial scale, there must be additional issues leading to both the location of the peak and the shape of the distribution with the degraded data. 

To explore this further, we looked closer at the detected BPs across the data sets to see any commonalities that could explain some of the variations. One of the most common issues leading to larger BPs was a direct consequence of the spatial resolution, in that features close to each other were identified as one single entity as opposed to separate features. This is not only a case of smaller features being missed, as one might expect, however, it is a scenario in which the size of BPs that should be easily distinguished in the degraded data is being conflated to include neighboring pixels of smaller features.

\begin{figure*}
    \centering
    \includegraphics[width=\textwidth]{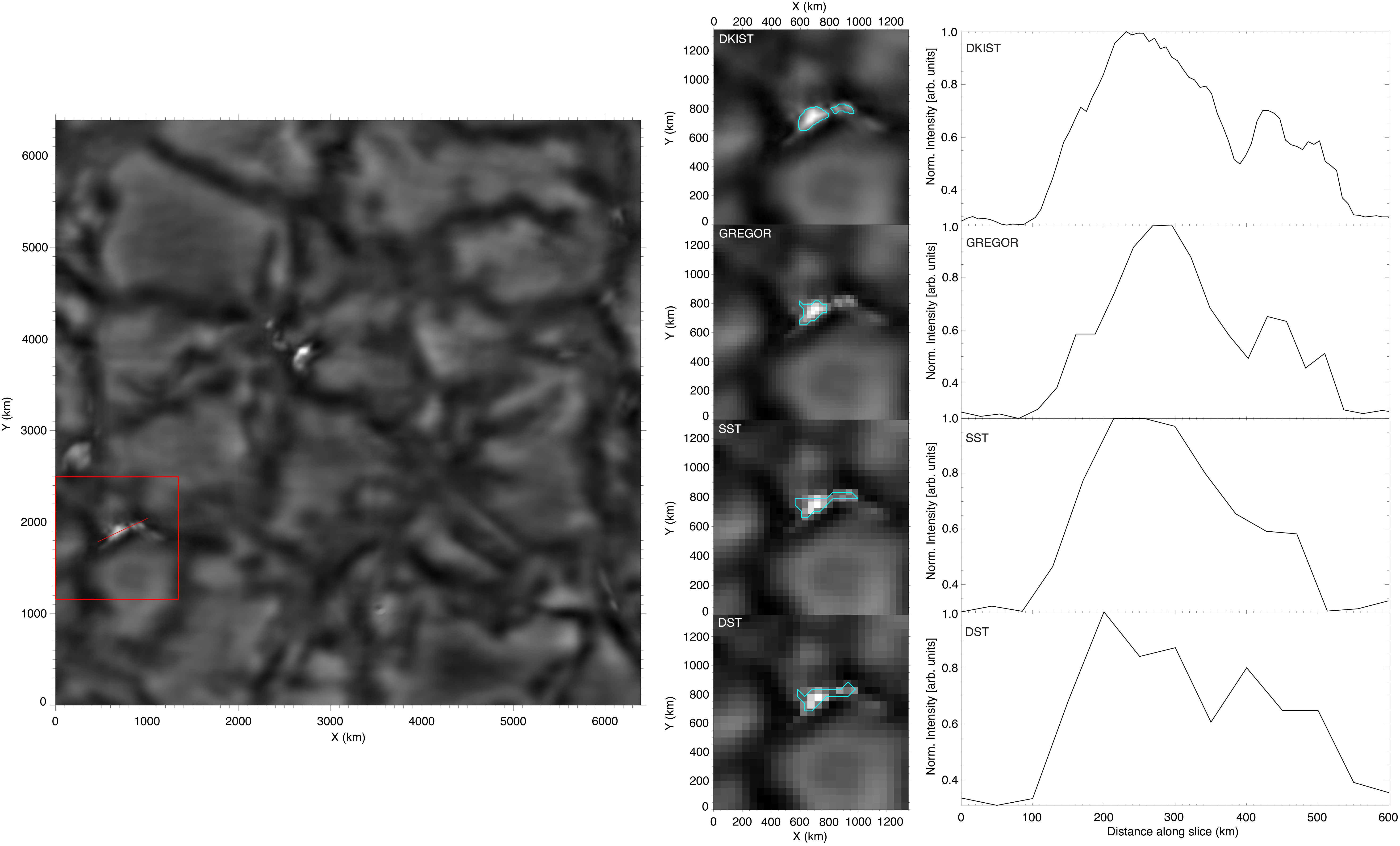}
    \caption{Sample images showing the determination of BP boundaries across the different data sets. The image on the left hand side shows a slightly expanded field-of-view of the region for context. The \textit{red} box indicates the region displayed in the middle panels, while the \textit{red} line across the BP indicates the position of the cross-cut used in the intensity plots in the far right panels. The middle panels indicate the appearance of the BPs across the different datasets, from top to bottom, going from best to worst spatial resolution. The {\textit{cyan}} colored contours indicate the detected boundary location of the BPs. The panels to the right of these intensity images give a sample light curve from the cross cut of the BP at the various resolutions. It should be noted that in this case, for DKIST data, the code identified two independent BPs, for GREGOR data, the code identified the larger BP but missed the smaller BP, and both BPs are identified as a single entity with the SST and DST data.}
    \label{degraded_inten_LC_BP_sample}
\end{figure*}

An example of this can be seen in Figure~\ref{degraded_inten_LC_BP_sample}. In this figure, a region is identified within the {\textit{red}} box of BPs in close proximity to each other. A {\textit{red}} line indicates one of the intensity cross-cuts used to identify the BP boundary locations. This direction was chosen to illustrate the effects of intensity thresholding at various spatial resolutions. The middle panels of the figure shows the BPs identified in each of the data sets with the {\textit{cyan}} contours indicating the boundary identified by the algorithm for that particular data set. The final panels on the right show the corresponding intensity curves used by the algorithm to identify the BP boundaries for the {\textit{red}} line shown in the far left panel. 

Within this figure, it should be clear that the algorithm identifies the same feature differently depending on the spatial resolution. For the DKIST data, two features are identified: a larger feature within the middle of the panel, with a smaller feature identified to the top right of the larger feature. For the GREGOR data, the larger feature is identified, but the smaller feature to the top right is missed. For both the SST and DST data sets, the larger feature and the smaller feature to the top right are identified as one individual feature. 

The intensity curves for each data set show how the algorithm comes to these conclusions. The algorithm looks for turning points at the edge of the BPs as the point at which there is a transition from brighter BP pixels to the darker surrounding intergranular lane pixels. To distinguish BPs from other small bright features, the algorithm uses a gradient threshold to separate real features from possible false detections. BPs will have a steeper intensity gradient going from the intergranular lanes to the center of the BP as the intensity at the center of the BP is above the quiescent level of the surrounding granules. 

For both the SST and DST data in the example shown in Figure~\ref{degraded_inten_LC_BP_sample}, the reduced spatial resolution results in the intensity in the region between the two features being `blended'. This acts to reduce the intensity gradient between the two features, so the algorithm does not detect them as individual features. As such, a single entity is identified, increasing the total area of that particular BP. Furthermore, for this particular case, the smaller feature to the top right has a double peak in intensity, albeit not sufficient to be identified as separate features in any data set. An effect of the reduction in spatial resolution here is that this double peak is not seen in the SST and DST data, though it is observed in both the DKIST and GREGOR data. 

For the GREGOR data, there is sufficient resolution to detect the boundary of the larger feature as the algorithm can pick out the turning point around 400\,km across the slice. However, for the smaller feature to the top right, there are insufficient pixels for the code to identify the left side of this smaller feature. As the algorithm cannot establish the boundary here, the BP is not identified. Hence, for the GREGOR data the larger feature has a more accurately determined boundary and, therefore, area component. Nevertheless, the smaller feature is not detected, and so smaller components of the BP area distribution are missed. 

This is the most common variation in boundary determination across the data sets. Smaller features are either not resolved at all at lower resolutions or not detected as the spatial resolution effectively `blurs' the boundaries between BP pixels and the intergranular lanes, and, therefore, the gradient of features is insufficient to identify a BP. As outlined above, numerous cases are seen in which nearby features are conflated to larger features at lower spatial resolution. 

An additional scenario that affects the BP area distribution occurs for larger scale features. This predominantly affects larger features identified in DKIST data and possibly explains the sharp tail in the DKIST distribution at larger area values. For larger BPs, like those found in BP chains, the algorithm occasionally misses longer BPs entirely as it considers them as false detections. This scenario occurs for BPs whereby the center of the BP is not sufficiently greater in intensity than the quiescent levels (i.e., only marginally brighter than granules). This results in the gradient in certain directions across a BP not being sufficient for the code to identify it as a feature. Instead, the algorithm treats it similarly to an exploding granule and discards the feature. 

Another issue with longer BPs occurs with the DKIST data as the algorithm segments longer chained features into smaller components, whereas, for lower resolution data sets, these chains are identified as singular features. This again acts to increase the number of smaller features detected in DKIST data. This scenario can be slightly more challenging to accurately interpret and essentially comes down to a somewhat subjective definition of what a BP is and where their boundaries lie. That is, the interpretation that these chains are made of individual features that are close enough together to appear as one long feature could be justified. As such, with the resolution of DKIST the algorithm detects these as individual smaller features. However, the intensity of the pixels between these individual bright centers does not return to levels similar to intergranular lane pixels. This suggests that there is still significant levels of flux between the bright centers. Therefore, it could be argued that these features are one entity with multiple bright centers representing regions of greater magnetic flux. This is clearly distinct from the scenario seen in Figure~\ref{degraded_inten_LC_BP_sample} with two features close together as the intensity drops to close to the level of the intergranular lanes in this scenario, which would be indicative of distinct features. Ultimately, this comes down to how the boundary of the BP is defined. Further work is needed here to explicitly determine the internal magnetic structures of these longer chains that have been segmented to support the discussion of whether they can be treated as individual features or one single chain. Unfortunately, there is insufficient data to determine the internal magnetic fields associated with these chains for these observations, and high spatial and temporal resolution 2D polarimetric data would be needed to further analyze these chains. 

 The issue with longer elements may be addressed by better techniques, e.g., deep learning techniques \citep{Yang2019, Xu2021}. However, these techniques are often supervised deep learning methods, and so require a large volume of annotated data \citep{Bai2023}, therefore, the issue of the interpretation of groups and chains may still pose a challenge. To reiterate, care must be taken when interpreting longer elements to determine whether these features should be interpreted as individual features or a conglomeration of several features. A suitable observing sequence with high spatial resolution polarimetric data would be needed to confirm which interpretation is correct. 

Another possible explanation for the variations observed between the DKIST results and the degraded data sets could come from the work of \citet{BerriosSaavedra2022}. Here, the authors analyzed BPs in GREGOR and Hinode datasets and found the usual log-normal distribution for BP areas. However, the authors report that the log-normal distribution is composed of two separate components, which were interpreted as two different populations of BPs from the network and intranetwork flux elements. Interestingly, for their GREGOR dataset the authors report a mean BP area of $5700$~km$^2$ for the lower area component and a mean of $16{\,}000$~km$^2$ for the larger area component. The DKIST distribution may be sampling this lower area component. It should be noted that the field-of-view of VBI is slightly larger than the typical widths of a supergranular cell \citep{Roudier2014, Noori2019}. It is possible then that this particular observed region preferentially samples the center of a network cell, and so, our sample of BPs is predominantly intranetwork BPs that produce this lower area component.

\subsection{Spatial Resolution}

One issue that could affect the area distribution reported for the DKIST data is due to variable seeing conditions across the data set. Seeing obviously is an issue with ground-based observations and can cause issues for small-scale features such as BPs, as they are close to the resolution limit. Seeing conditions were relatively good across this data set, although there were some fluctuations (Figure~\ref{spat_res_var}). 

When accounting for seeing conditions by filtering detections below the smallest resolvable pixel dimensions in a given frame, the peak of the area distribution for the DKIST BPs was revised to $4800$\,km$^2$. This is similar to the values for the lower area components reported by \citet{BerriosSaavedra2022}. The peak for the filtered area is approximately 2.1 times larger than that using the spatial sampling of the data. The distribution can again be fitted well with a log-normal distribution. The tail of the distribution is also less sharp than the original distribution.

It is clear that care must be taken when analyzing spatial characteristics of small-scale elements such as BPs, particularly for data susceptible to seeing fluctuations. Small fluctuations can have a large impact on small features \citep{Abramenko2010, Riethmuller2014, BerriosSaavedra2022}. Looking at the adjusted distribution, it is clear that the distribution of area values is larger than initially estimated, although still lower than the values estimated with facilities with lower spatial resolutions. This distribution is likely to still estimate lower values for areas due to the issues mentioned in Section~\ref{sect:disareas}, that is, longer chains are misinterpreted as false detections and excluded, as well as longer elements being segmented by the algorithm into smaller components. However, the peak is still consistent with the lower component reported by \citet{BerriosSaavedra2022}. This suggests that, even after adjusting for the effects of seeing, the superior resolving power of DKIST coupled with the field-of-view of VBI means that smaller intranetwork BPs are adequately resolved and preferentially sampled with these observations. 

The results of \citet{Kuridze2025} are relevant when considering the area distribution and size of the BPs. Here, the authors utilized the field sampling mode of the VBI to create a mosaic (with $3 \times 3$ tiles) of a larger field-of-view of the photosphere, at the cost of temporal resolution. The authors image faculae off disc center ($\mu \approx 0.85$) and report spatial resolutions in the range $0{\,}.{\!\!}{''}022$\,--\,$0{\,}.{\!\!}{''}030$, which was determined based on the smallest resolved structures observed within the data. Compared with MURaM simulations \citep{Rempel2014}, the authors interpret striations near granular edges as due to the magnetic flux density within the lanes creating a Wilson depression and report widths in the range of $20$\,--\,$50$~km for the eight striae that were sampled. 

Faculae can be interpreted as BPs viewed at an angle \citep{Carlsson2004}. Therefore, the results with regards to the width of the striations found by \citet{Kuridze2025} appear consistent with the results we present here for BP diameters (particularly with our original unfiltered results of $54$~km) at the upper range of their reported widths. More details are needed to evaluate their estimation of the spatial resolution, as there are insufficient details on how this was performed. There is a potential selection bias introduced when considering the smallest resolved structures, which the Fourier technique should avoid. This, coupled with an average r$_0$ of $12.2$~cm for this data (which is comparable to the data presented here), suggests that the spatial resolution of their data could be larger than reported. Nevertheless, the MURaM simulations presented in the work of \citet{Kuridze2025} supports striations of these widths and possibly smaller. This highlights the challenges in interpreting the scale of these features and the role of simulations in supporting observational work.

Regardless, the results with the area distributions confirm that smaller-scale elements than previously reported are observable with DKIST, and so DKIST data are necessary to accurately classify this smaller component of the BP distribution. This is further highlighted when limiting the area distribution to the 6 best frames in the data set, which resulted in a distribution similar to our original unfiltered area distribution, suggesting that these smaller features are more readily identified in DKIST data. Previous work looking at area distributions likely had systematic errors in the reported statistics as a result of the resolving power. However, the results indicate that facilities such as GREGOR should be able to adequately sample these smaller scale features.


\subsection{Bright Point Dynamics}
Velocity characteristics were largely similar between the DKIST data and the DKIST data degraded to the resolution of GREGOR, the SST and the DST. The peak in velocity occurs at $0.83$\,km\,s$^{-1}$ for DKIST and $1.25$\,km\,s$^{-1}$ for the degraded data sets, as seen in Figure~\ref{Fig_vel_char}. The degraded data sets have a longer tail of higher velocity BPs. This results in a higher mean BP velocity for the degraded data ($2.30\pm0.64$\,km\,s$^{-1}$, $2.44\pm0.68$\,km\,s$^{-1}$, and $2.36\pm0.66$\,km\,s$^{-1}$ for the GREGOR, SST and DST data, respectively) in comparison to the DKIST data ($1.60\pm0.41$\,km\,s$^{-1}$). These values are in agreement with results from the literature \citep[e.g., with peaks found around $1$\,--\,$2$\,km\,s$^{-1}$ in studies by][to name a few]{Utz2010, Keys2011, Chitta2012}.

The difference in velocity characteristics at higher velocities is partially due to similar issues as outlined in Section~\ref{sect:disareas}, whereby BPs in close proximity to each other can be detected as one element when the spatial resolution is lower (see Figure~\ref{degraded_inten_LC_BP_sample}). In these scenarios, it is possible that between frames the BPs in close proximity can separate enough so that they are detected as separate entities, which shifts the barycenter of the original BP. Similarly, if two separate entities are close enough to each other that in a subsequent frame they are identified as a single object, this can be enough to move the barycenter location by a few pixels, which will be registered as a large transversal velocity excursion, which acts to artificially broaden the velocity distribution at higher velocity values. These issues will be exacerbated as the reduced spatial resolution of the degraded data sets will have a higher degree of uncertainty associated with the spatial dimensions of the pixels. This is further compounded by variable seeing conditions, whereby a drop in image quality due to atmospheric variations may be enough to `smear' the boundary of two features in close proximity, which will then be identified as one in the frame affected by poorer seeing conditions. 

Figure~\ref{corrected_vel} shows the effects on the velocity distribution when adjusted for the effects of seeing. Filtering the data to remove features that are not adequately resolved due to seeing variations, the peak of the distribution shifts to a slightly higher velocity ($\sim 1.18$\,km\,s$^{-1}$) and is now coincident with the peaks for the degraded data sets. The shape of the DKIST distribution broadens with this adjustment, with the mean velocity now $2.28 \pm 0.58$~km\,s$^{-1}$. This can be interpreted as the seeing-induced reduction in spatial resolution affecting the sample of BPs resolvable. This results in larger BPs within the sample, which are susceptible to errors in transverse velocity estimates due to barycenter fluctuations. This results in a shift of the peak in the velocity distribution to higher velocities and a broadening of the distribution with more higher velocity BPs reported. The true velocity distribution of BPs may be lower than those in the literature (similar to the the unfiltered DKIST results); however, data with more consistent seeing conditions throughout, with a resolution close to the diffraction limit of DKIST are needed to confirm. With the present data, the velocity characteristics appear consistent with previous work.

The value for $\gamma$ was estimated as $1.22\pm0.41$ for the DKIST data ($1.24\pm0.43$ for the seeing filtered DKIST data) and was estimated as $1.40\pm0.58$, $1.52\pm0.69$, $1.55\pm0.73$ for the GREGOR, SST and DST data, respectively. This implies that the BPs within the DKIST data are super-diffusive. For a BP's transverse motion, this appears as a random walk pattern, with occasional L{\'{e}}vy flights. This is consistent with previous work on the diffusion of BPs \citep{Giannattasio2014, Keys2014, Yang2015, Giannattasio2019}. 

The higher super-diffusive element of motions in the degraded data sets is likely due to shifts in barycenter positions for cases of BPs in close proximity identified as one feature in one frame and as separate in the next. This is exacerbated at lower spatial resolutions, as these features may be identified erroneously as one entity. This is reflected in the fact that the $\gamma$ value is higher at lower resolutions (i.e., for the degraded data sets), which implies more L{\'{e}}vy flights for BPs in the lower resolution data sets. Previous work with DST data \citep{Keys2014} found a $\gamma$ value of $\sim1.2$, so it is possible that the BPs in this particular observation are slightly more super-diffusive than previously observed. This is supported by the fact that the velocity distribution is more elevated than the result of \citet{Keys2014} (who found mean velocities in the range $\sim 0.6$\,--\,$0.9$~km\,s$^{-1}$).

Values for $K$ were estimated as $30\pm18$~km$^2$\,s$^{-1}$ for the DKIST data ($74\pm47$~km$^2$\,s$^{-1}$ for the seeing filtered DKIST data) and was estimated as $98\pm55$~km$^2$\,s$^{-1}$, $142\pm83$~km$^2$\,s$^{-1}$ and $144\pm83$~km$^2$\,s$^{-1}$ for the GREGOR, SST and DST data, respectively. The values here are consistent with those found in the literature \citep{Berger1998,Hagenaar1999, Abramenko2011, Jafarzadeh2017, BellotRubio2019, Rincon2025}. The value for the original unfiltered DKIST data is somewhat lower than the other data sets. Values for $K$ for BPs typically are below $\sim300$~km$^2$\,s$^{-1}$, with many values reported below $\sim100$~km$^2$\,s$^{-1}$ \citep{Cameron2011, Chitta2012, Yang2015}. The value for $K$ has been reported as low as $12$~km$^2$\,s$^{-1}$ \citep{Abramenko2011}. The lower values obtained with BPs is due to the fact that their motion and transport is governed by the shorter-lived, smaller-scale structures such as granules as opposed to supergranular cells \citep{Rincon2025}. 

The variation between the values of $K$ observed for the DKIST data in comparison to the degraded data is not unexpected. It has been shown for super diffusive motions ($\gamma > 1$), $K$ is directly proportional to both the temporal and spatial scales \citep{Abramenko2011}, with the value decreasing with decreasing temporal and spatial scales. Therefore, as the spatial resolution decreases, the value of $K$ will increase, which is what is reported here. 

The average lifetimes were estimated as $95\pm29$~s, $97\pm32$~s, $98\pm33$~s, and $97\pm33$~s for the DKIST, GREGOR, SST and DST data, respectively. The values correspond to values obtained in previous studies on quiet Sun BPs \citep{Keys2011,Keys2014}. This suggests that around 1.5~minutes is a typical average lifetime of BPs in the quiet Sun. Lifetimes may be larger on average for network BPs or BPs in the vicinity of an active region. This difference between isolated and non-isolated BP lifetimes has been reported by \citet{Liu2018}, with non-isolated BPs surviving on average around 1.5 times longer than isolated BPs. Future work will look to establish the variations due to the location within the network cell. Again, it should be noted that the field-of-view of VBI is slightly larger than the typical widths of a supergranular cell, so additional data would be required to classify BPs as network or internetwork. However, the average lifetimes we report are consistent across the DKIST data and the degraded data sets and follow those in the literature. 

\section{Conclusions}\label{sect:conc}
Data from the newly commissioned DKIST were utilized to determine the characteristics of small-scale features (BPs) across the solar surface at the highest spatial resolutions currently achievable. This is the first analysis of the properties of BPs with DKIST, and the highest resolution analysis of BPs ever performed. 

The VBI instrument was used to obtain photospheric images of a quiet Sun region at disc center with the G-band filter for around 90~minutes using the highest spatial sampling possible ($0{\,}.{\!\!}{''}011$~pixel$^{-1}$) with DKIST. The analysis of the images showed that the average spatial resolution of the data was around $0{\,}.{\!\!}{''}034\pm0{\,}.{\!\!}{''}007$, with a handful of frames obtained close to the diffraction-limit with a spatial resolution of $0{\,}.{\!\!}{''}023$. The lowest quality image had an estimated spatial resolution of $0{\,}.{\!\!}{''}066$ within this data set. The data has an average r$_0$ value of $12.1\pm 3.1$~cm and a range of $10.2$~cm to $20.9$~cm.

BPs are small-scale intensity enhancements that are often the manifestation of kilogauss magnetic fields observed between granules on the solar surface. An automated detection and tracking algorithm was employed to identify BPs across the data set so that the properties could be analyzed. BPs were observed to have an average lifetime of $95\pm29$~s, which is consistent with previous studies. The BPs were found to have a log-normal velocity distribution with a peak at $0.83$\,km\,s$^{-1}$ and a mean velocity of $1.60\pm0.41$\,km\,s$^{-1}$. Again, this is consistent with previous work. 

The areas of the BPs were also established, which showed the biggest difference from previous work. The area of the BPs was found to have a log-normal distribution with a peak at $2300$\,km$^2$. This is somewhat lower than values found in previous studies \citep{Crockett2010, Keys2014}, although it is perhaps due to a lower area component seen in the literature \citep{BerriosSaavedra2022}. It is possible that the higher spatial resolution obtained with DKIST is able to resolve smaller features more readily. With this, the distribution was narrower than those seen previously, suggesting that smaller features were observed than in previous work, again highlighting that the lower area component was preferentially sampled. This is supported by the field-of-view size of VBI, which is close to the scale of a supergranular cell. It is possible then that intranetwork BPs are preferentially sampled here, returning the lower area component reported by \citep{BerriosSaavedra2022}.

To compare with previous results more readily, the DKIST data were degraded to match the resolution of commonly used ground-based facilities, namely, GREGOR, the SST and the DST. The same detection and tracking algorithm was applied to the degraded data sets. The degraded data sets showed the same log-normal distributions for the area as the DKIST data, albeit with peaks at higher areas. The distributions from the degraded data sets matched well with area distributions reported previously for these facilities. It was observed that for the data sets with lower spatial resolution, frequently smaller groupings of BPs would be identified as a single entity as there was insufficient resolving power to determine accurate boundaries. This acted to increase the number of larger features observed in these data. To obtain a more accurate picture of BP area distributions, the area estimates were adjusted for the seeing-induced reduction in spatial resolution in the images for the DKIST data by filtering out BPs that fell below the resolution limit in a given frame. This led to the area distribution shifting to slightly higher values, with the peak now found at $4800$\,km$^2$. The peak still falls well below that of the degraded examples, but is likely closer to the `true' distribution of the areas for this data set. Furthermore, isolating the area distribution for the 6 best frames in the data set returned a distribution similar to the original, unfiltered data set, suggesting that smaller features are more readily identified in DKIST data. It may be that a data set with more consistently excellent seeing throughout will exhibit this distribution with a peak at lower areas.

The observed area distribution accuracy will only be improved with data with more consistent seeing for the duration, so that the images are close to the diffraction-limit for the duration of the sequence. This would ensure that the BPs are sampled with consistently excellent spatial resolution, which is vitally important for features at this scale. If consistent quality across a sequence could be achieved, the limiting factor for resolving BPs for a telescope with the aperture width of DKIST, would be the mean free path of a photon in the photosphere. Other techniques may be needed to interpret the structuring of longer-chains of BPs, which are possibly underestimated with the current detection algorithm employed with DKIST data.

The velocity and lifetimes of the BPs were also analyzed with the degraded data sets. The average lifetimes of the BPs were consistent with the DKIST results and with previous studies. The velocity distributions for the degraded data sets was consistent with previous studies and slightly elevated in comparison to the results from DKIST. The difference was due to the spatial resolution, whereby there are larger errors in the estimation of the transverse velocity when the spatial resolution is reduced. Similarly, some BPs had high velocities due to multiple features in close proximity to each other. These BPs were first identified separately and then as a single entity between frames due to inadequate resolving power for the features. This meant that the barycenter location of the BP could shift significantly between frames due to the change in size and shape of the larger BP, which then translated into a higher velocity estimate for these features. Therefore, care should be taken when considering BPs in groups (e.g., around a network boundary) when establishing their dynamic characteristics.

Features in close proximity can be an issue when dealing with feature tracking and arise from the feature classification process. That is, how an algorithm defines the boundary of features and how individual features are then classified between frames to identify the continuation of features will never be completely accurate. Edge cases, such as features being identified as separate then as one between subsequent frames, will add to the inaccuracy of the algorithm. 

Similar issues exist for longer chains of BPs, whereby the constraints of the algorithm may over segment features. This can be a challenging identification task, as it can be open to interpretation and potential biases, for example, if dips in intensity across a chain are indicative of internal structuring of a singular entity or if they are markers of feature boundaries for features in close proximity. Such image classification issues will exist with machine learning techniques as well \citep{Bai2023}, so other techniques and data may likely be necessary to understand the complex structuring of BP chains and closely packed groups. 

Future work will look at signatures of the BPs in other filters and will make use of alternative techniques to understand the underlying structure of BPs. This will aid our understanding of the magnetic field at the smallest scales and its coupling to higher regions of the solar atmosphere.

\begin{acknowledgments}
The authors are grateful to the anonymous referee for suggestions to improve the manuscript. RJC, MM and DBJ acknowledge support from the Science and Technology Facilities Council (STFC) under grant No. ST$/$P000304$/$1 $, $ ST$/$T00021X$/$1 \& ST$/$X000923$/$1. DKJM acknowledges a studentship funded by the Leverhulme Interdisciplinary Network on Algorithmic Solutions. RE acknowledges the NKFIH (OTKA, grant No. K142987) Hungary for enabling this research. RE is also grateful to the Science and Technology Facilities Council (STFC, grant No. ST/M000826/1) UK, PIFI (China, grant No. 2024PVA0043), and the NKFIH Excellence Grant TKP2021-NKTA-64 (Hungary). DJC acknowledges partial support of this project from NASA grants 19-HSODS-004 and 21-SMDSS21-0047. The research reported herein is based on data collected with the Daniel K. Inouye Solar Telescope (DKIST), a facility of the National Solar Observatory (NSO). DKIST is located on land of spiritual and cultural significance to Native Hawaiian people. The use of this important site to further scientific knowledge is done so with appreciation and respect. NSO is managed by the Association of Universities for Research in Astronomy, Inc., and is funded by the National Science Foundation. Any opinions, findings and conclusions or recommendations expressed in this publication are those of the author(s) and do not necessarily reflect the views of the National Science Foundation or the Association of Universities for Research in Astronomy, Inc.  The observational data used during this research is openly available. Data availability statement: readers can access the DKIST data used in this study from the \hyperlink{https://dkist.data.nso.edu}{DKIST Data center Archive} under proposal identifier pid$\_1\_36$. Finally, we wish to acknowledge scientific discussions with the Waves in the Lower Solar Atmosphere (WaLSA; \href{https://www.WaLSA.team}{https://www.WaLSA.team}) team, which has been supported by the Research Council of Norway (project no. 262622), The Royal Society \citep[award no. Hooke18b/SCTM;][]{2021RSPTA.37900169J}, and the International Space Science Institute (ISSI Team~502).  
\end{acknowledgments}


%

\textbf{\facilities{DKIST \citep{DKIST2020}.}}



\bibliography{dkist_mbps_biblio}{}
\bibliographystyle{aasjournal}

\end{document}